\documentclass[sigconf]{acmart}
\usepackage{booktabs} % For formal tables
\usepackage[font=small]{caption}
\usepackage[labelformat=simple]{subcaption}
\usepackage{graphicx}
\usepackage[ruled,vlined,linesnumbered]{algorithm2e}
\usepackage{algorithmic}
\SetKwComment{Comment}{$\triangleright$\ }{}
\usepackage[mathscr]{eucal}
\usepackage{color}
\usepackage{url}
\usepackage{mathtools}
\usepackage{xcolor}
\usepackage{multirow}
\usepackage{soul}
\usepackage{colortbl}
\usepackage[nodisplayskipstretch]{setspace}

\renewcommand{\textrightarrow}{$\rightarrow$}

\allowdisplaybreaks

\newcommand{\method}{\textsc{ISCON}\xspace}
\newcommand{\grurec}{\textsc{GRU4REC}\xspace}
\newcommand{\tagnn}{\textsc{TAGNN}\xspace}
\newcommand{\cotrec}{\textsc{COTREC}\xspace}

\newcommand{\csrm}{\textsc{CSRM}\xspace}

\copyrightyear{2022}
\acmYear{2022}
\setcopyright{acmcopyright}\acmConference[CIKM '22]{Proceedings of the 31st ACM
International Conference on Information and Knowledge Management}{October
17--21, 2022}{Atlanta, GA, USA}
\acmBooktitle{Proceedings of the 31st ACM International Conference on Information
and Knowledge Management (CIKM '22), October 17--21, 2022, Atlanta, GA, USA}
\acmPrice{15.00}
\acmDOI{10.1145/3511808.3557613}
\acmISBN{978-1-4503-9236-5/22/10}



\begin{document}

\title{Implicit Session Contexts for Next-Item Recommendations}

\author{Sejoon Oh}
\email{soh337@gatech.edu} 
\affiliation{%
  \institution{Georgia Institute of Technology}
    \country{United States}
}

\author{Ankur Bhardwaj}
\email{ankurbhardwaj843@gmail.com} 
\affiliation{%
  \institution{Georgia Institute of Technology}
    \country{United States}
}

\author{Jongseok Han}
\email{jhan405@gatech.edu} 
\affiliation{%
  \institution{Georgia Institute of Technology}
    \country{United States}
}

\author{Sungchul Kim}
\email{sukim@adobe.com}
\affiliation{%
  \institution{Adobe Research}
    \country{United States}
}

\author{Ryan A. Rossi}
\email{ryrossi@adobe.com}
\affiliation{%
  \institution{Adobe Research}
    \country{United States}
}

\author{Srijan Kumar}
\email{srijan@gatech.edu}
\affiliation{%
  \institution{Georgia Institute of Technology}
    \country{United States}
}

\renewcommand{\shortauthors}{Oh, et al.}

\begin{abstract}
\noindent Session-based recommender systems capture the short-term interest of a user within a session. 
Session contexts (i.e., a user's high-level interests or intents within a session) are not explicitly given in most datasets, and implicitly inferring session context as an aggregation of item-level attributes is crude. 
In this paper, we propose  \method, which implicitly contextualizes sessions. 
\method first generates implicit contexts for sessions by creating a session-item graph, learning graph embeddings, and clustering to assign sessions to contexts. 
\method then trains a session context predictor and uses the predicted contexts' embeddings to enhance the next-item prediction accuracy.
Experiments on four datasets show that \method has superior next-item prediction accuracy than state-of-the-art models. 
A case study of \method on the Reddit dataset confirms that assigned session contexts are unique and meaningful. 
\end{abstract}

% and recommend relevant items according to the current interest of the user

\maketitle

\section{\textbf{Introduction}}
\label{sec:intro}
Session-based recommendation systems (SBRSs)~\cite{hansen2020contextual, wu2019session, liu2018stamp, yu2020tagnn, COTREC, hidasi2018recurrent, you2019hierarchical, li2017neural, wang2019collaborative, pan2020intent, ren2019repeatnet} have been proposed to accurately model a user's short-term and evolving interest, where a user's session is defined as a sequence of its interactions with items occurring within a short time period~\cite{wang2021survey,  hidasi2018recurrent}. 
SBRSs incorporate the current interest of a user within a session as well as the evolving preferences of a user across consecutive sessions.

\textit{Context-aware} recommendations~\cite{Ltcross, hansen2020contextual, twardowski2016modelling, sheu2020context, gabriel2019contextual, wang2020make, pan2020intent, wang2019collaborative, wang2020global} have also gained attention as contexts can represent high-quality knowledge about the user's interests, which can enhance next-item prediction performance. 
In this paper, we define the contexts of a session as a user's high-level interests and objectives during the session. 
Our goal in the paper is to predict session context, as it can serve as novel features to enhance the next-item prediction quality in the current session. 
Furthermore, since consecutive sessions are likely to have related user interests (i.e., complementary, similar, or supplementary)~\cite{li2017neural}, the current session context can help make better recommendations in the next session, especially when the next session only has a few interactions. 

Existing context-aware models have limitations in SBRSs that (1) they require session context information to be given explicitly, but in reality, the session contexts are often unavailable in the data or may need to be implicitly inferred, (2) they cannot incorporate session contexts to their models, (3) they employ inaccurate implicit session contexts when only the first few items of a session are observed, and (4) simple aggregation of item features to derive implicit session contexts can be inaccurate.
Since session contexts are rarely presented explicitly, it is crucial to create methods that can assign precise and meaningful implicit contexts to sessions.  

We propose a novel recommendation model called \method (\underline{I}mplicit \underline{S}ession \underline{CON}texts for next-item recommendations). \method first finds implicit contexts of sessions via graph-based session contextualization, which is more meaningful and precise compared to existing session contextualizations using simple item feature aggregation. \method trains a context predictor using the implicit contexts as labels to predict future sessions' contexts accurately. Finally, \method utilizes the predicted contexts as novel features to enhance the next-item prediction accuracy. 
The main novelties of \method include that (1) \method develops an \textit{\textbf{implicit session context predictor}} that estimates session contexts (even for sessions with few items and for future sessions), (2) a \textit{\textbf{next-item predictor that leverages predicted session contexts}} and merges them with other features.
Experimentally, \method outperforms 4 state-of-the-art SBRSs across 4 real-world datasets. Moreover, we perform an ablation study of \method to confirm the effectiveness of session contextualizations (in Appendix). 
A case study of \method on the Reddit dataset shows that the sessions are properly contextualized.
Our dataset and code used in the paper are available here\footnote{\url{https://github.com/srijankr/iscon}}.

\section{\textbf{Related Work}}
\label{sec:related_work}
Context-aware recommender systems~\cite{Ltcross, hansen2020contextual, sheu2020context, gabriel2019contextual, xu2019graph, yuan2020future, wang2020make, pan2020intent, wang2019collaborative, wang2020global} incorporate contextual information into their models for capturing user preferences correctly~\cite{KULKARNI2020100255}.
A session context can imply various aspects such as temporal features~\cite{Ltcross,wang2020make, ma2020temporal} and graph-based features~\cite{xu2019graph, wang2020make, sheu2020context}; our context definition is high-level intents or interests of a user in a session, which is related to multi-interest extraction methods~\cite{tan2021sparse, li2019multi, cen2020controllable}.
Many algorithms assume the context information is given~\cite{Ltcross, hansen2020contextual, gabriel2019contextual, hu2017diversifying, wang2021survey} or perform user- or interaction-level contextualization~\cite{xu2019graph, yuan2020future, tan2021sparse, li2019multi, cen2020controllable}, not session-level.
Few methods~\cite{pan2020intent, wang2019collaborative, xu2019graph, wang2020global} are able to contextualize sessions, but their contextualizations can be inaccurate for sessions with few items or sessions that are not included in the training, since they do not utilize other session information or cannot inductively contextualize sessions without model retraining (e.g., In Table~\ref{tab:next_item_accuracy}, \method outperforms \csrm that also contextualizes sessions).

For SBRSs, attention mechanisms have been widely adopted~\cite{li2017neural,liu2018stamp, wang2019collaborative, pan2020intent, xu2019graph, ren2019repeatnet}.
Graph neural network-based models~\cite{yu2020tagnn, COTREC, wu2019session, xu2019graph, wang2020global} also achieve superior performance by capturing complex transitions of items on a session-item graph.
These models cannot assign implicit contexts to sessions accurately, predict session contexts, or use the session contexts for the next-item prediction.

\section{Proposed Approach: \method}
\label{sec:proposed_method}
\subsection{Contextualizing Sessions}
\label{sec:method:context_assignment}
We define the context of a session as a summary of interests  expressed by a user's interactions in a session.
A session's contexts can be used as prior knowledge to predict items that the user is likely to be interested in the session, and this can enhance the next-item prediction accuracy.
Most public datasets do not include explicit contexts (i.e., interests specifically stated by a user) of a session as it is hard to gather; for example, it is intrusive and disruptive to ask users about their current session contexts/interests directly. 
Unlike explicit contexts, implicit contexts of a session can be inferred interests from the users' interactions. Trivial approaches such as aggregating item features in a session cannot find proper implicit contexts when only the first few items of the session are observed or item features are uniformly distributed, which shows the need for a more sophisticated session contextualization method.

Our session contextualization approach consists of two steps: 1) generating session embeddings from a user-item multigraph, and 2) clustering sessions to identify session context clusters. To obtain session embeddings, we create a session-item bipartite multigraph, where its nodes are sessions and items, and its edges (undirected) represent membership between an item and a session. 
We apply a node embedding method called GraphSage~\cite{hamilton2017inductive} to the bipartite multigraph to obtain session embeddings. GraphSage~\cite{hamilton2017inductive} is used as its inductive capability that does not require retraining when computing the embeddings of future or test sessions.
The key advantages of this graph-based technique include (1) it does not require pre-trained item embeddings, and (2) by considering multi-hop paths in the graph, this method can produce generalizable session embeddings that encode not only the items in the current session but also those in nearby sessions in the graph. 
Next, given the number of implicit session contexts $|C|$ (hyperparameter; found by empirical searches), we cluster sessions by applying the K-means clustering~\cite{kanungo2002efficient} ($K=|C|$) to session embeddings. Each cluster represents similar sessions with the same implicit context. We define an implicit context of each session as  the number of a cluster closest to the session embedding. We define trainable embeddings of all $|C|$ implicit session contexts as session context embeddings.

\subsection{Session Context Prediction}
\label{sec:method:context_prediction}
Predicting the context of a future session and updating the context of the session in real-time (as new items are observed) can help guide the next-item prediction task. However, using the context assignment method in Section~\ref{sec:method:context_assignment} is not scalable, as multi-hop aggregation in the multigraph after observing every new item to re-generate session embeddings in real-time is expensive.
Moreover, this method does not consider the relationships across consecutive sessions of a user. Naturally, the contexts of consecutive sessions of a user can be related (complementary, similar, or supplementary)~\cite{li2017neural}. The previous sessions' context information can thus be used to predict the current/future session's context better.

\begin{figure}[t!]
    \centering
    \includegraphics[width=1.05\linewidth]{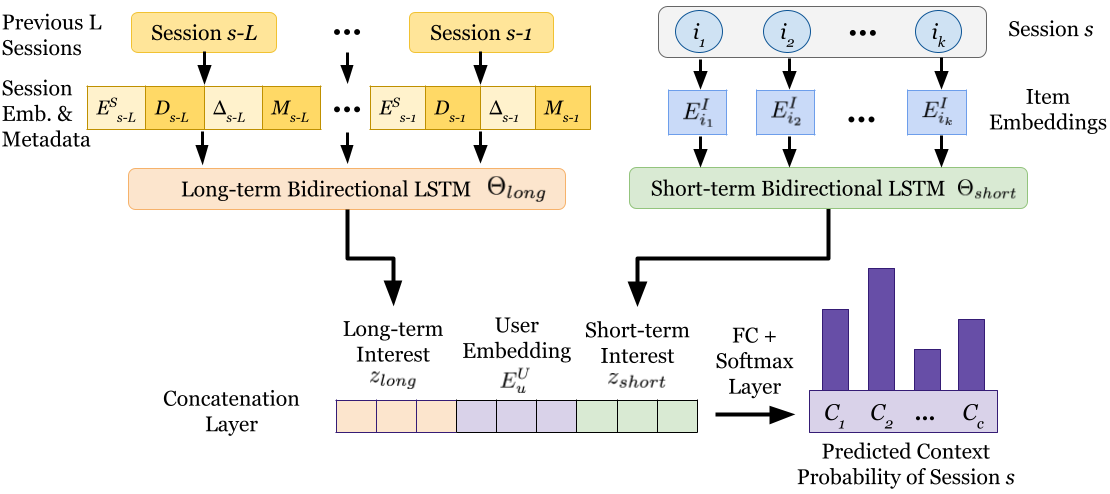}
    \caption{Context predictor of \method. 
    }
    \label{fig:context_prediction}
\end{figure}

We train a novel real-time session context predictor using the user's short-term (current session) and long-term (previous sessions) interest vectors as well as a trainable user embedding (see Figure~\ref{fig:context_prediction}). The predicted session context is updated dynamically whenever we observe a new item in a session.
We train two Bidirectional LSTMs (Bi-LSTMs)~\cite{schuster1997bidirectional} to get both vectors. 
Bi-LSTMs have shown superior performance in recommendation than LSTMs by utilizing both direction sequences~\cite{zhao2020deep, fu2018sequence}.
We have tried other architectures such as Transformer~\cite{vaswani2017attention} or GRU~\cite{cho2014properties} for \method, but they have empirically shown similar or worse prediction performance compared to the performance of the Bi-LSTM.

To derive the user's long-term interest vector using a \textit{long-term (session-level) Bi-LSTM}, we feed to it a sequence of previous $L$ sessions' features including the session embedding from Section~\ref{sec:method:context_assignment} and metadata. 
The metadata of a session $s$ includes the session duration $D_{s}$ (in seconds), the time interval $\Delta_{s}$ between sessions $s$ and $s+1$, and the number of items $M_s$ in the session. Given the current session $s$ of a user $u$, the output $z_{long}$ from the long-term Bi-LSTM $\Theta_{long}$ is given as follows:
\begin{equation}
    \begin{aligned}
z_{long} = \Theta_{long}([F_{s-L}, \ldots , F_{s-1}]) \\
F_{s} = concat(E_{s}^{S}, D_{s}, \Delta_{s},M_{s})
    \end{aligned}
\end{equation}
where $L$ is the maximum sequence length (a hyperparameter), and $E_{s}^{S}$ is the session embedding of the session $s$ (from Section~\ref{sec:method:context_assignment}).

To obtain the short-term interest representation of a user using a \textit{short-term (item-level) Bi-LSTM} $\Theta_{short}$, we input the sequence of embeddings of the observed items in the session.
The short-term vector is updated whenever we observe a new item in the session to make our context prediction more accurate.
Given the session $s$ of a user $u$ and observed items $i_1, \ldots , i_{k}$ in the session so far, the output $z_{short}$ from the short-term Bi-LSTM $\Theta_{short}$ is given as follows: 
\begin{equation}
    \begin{aligned}
z_{short} = \Theta_{short}([E^{I}_{i_1}, \ldots, E^{I}_{i_k}])
\end{aligned}
\end{equation}
where $E_{i}^{I}$ is the item embedding of an item $i$.~\footnote{
We use learnable item embeddings instead of using the node embeddings from Section~\ref{sec:method:context_assignment} since the trainable ones show higher next-item prediction accuracy empirically. 
}
We use an auxiliary vector (which is fixed for all sessions) or the last item's embedding of the user as input  when there are no observed items in a session.

Finally, we use a learnable user embedding $E^{U}_{u}$ as one of the input features for the context predictor. 
We expect improved personalization with $E^{U}_{u}$ since it is trained only with interactions of that particular user and is a representation of a user's overall behavior. 

We concatenate the three vectors -- long-term interest, short-term interest, and a user embedding -- and feed them to fully connected and Softmax layers. The output is the session context prediction vector $\hat{y} \in \mathbb{R}^{|C| \times 1}$, where $|C|$ is the number of session contexts. We note that contexts in Sections~\ref{sec:method:context_assignment} and \ref{sec:method:context_prediction} are the same, and we use all $|C|$ contexts derived in Section~\ref{sec:method:context_assignment} as candidates below.
\begin{equation}
    \begin{split}
\hat{z}_{pred-con} &= FC_{1}(concat(E^{U}_{u},z_{short},z_{long})) \\
\hat{y} &= softmax(\hat{z}_{pred-con})
\end{split}
\end{equation}

We train the context predictor $\Theta_{context} = \{\Theta_{short},\Theta_{long},$ $E^{U},E^{I},FC_1\}$ with the following Cross-Entropy loss and Adam optimizer on all training interactions $X_{train}$. For training in a supervised manner, we use the implicit contexts of the sessions derived in Section~\ref{sec:method:context_assignment} as ground-truth labels. The training loss is given as follows: 
\begin{equation}
    \begin{aligned}
        \mathbf{L}_{context}(\hat{y}) = -\sum_{i=1}^{|C|} {y_i \log (\hat{y_{i}})}
    \end{aligned}
    \label{eq:context_prediction}
\end{equation}
where $y$ is a one-hot vector containing the implicit context assignment of a current session $s$ of a user $u$ (\textbf{the output of Section~\ref{sec:method:context_assignment}}).

Using the trained session context prediction model $\Theta_{context}$, \method generates a session context probability vector ($\hat{y} \in \mathbb{R}^{|C| \times 1}$).

\subsection{Next-item Predictions with Session Contexts}
\label{sec:method:next_item_prediction}
\begin{figure}[t!]
    \centering
    \includegraphics[width=1.02\linewidth]{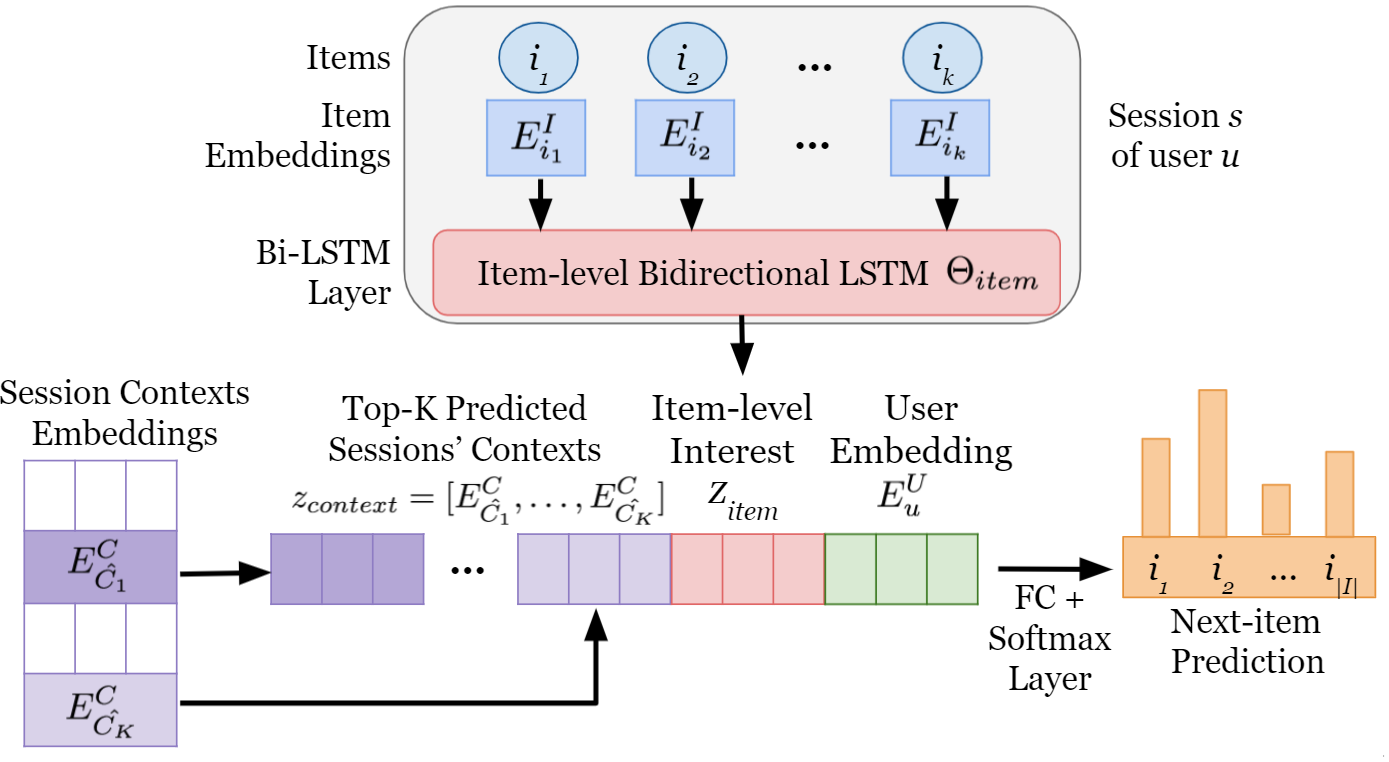}
    \caption{Next-item prediction with contextual embeddings.}
    \label{fig:next_item_prediction}
\end{figure}

The ultimate goal of \method is to enhance the next-item prediction accuracy by contextualizing sessions and utilizing those session contexts as indicators. 
Figure~\ref{fig:next_item_prediction} shows the \method architecture, where it combines predicted session context embeddings $z_{context}$, an item-level interest $z_{item}$, and a user embedding $E_{u}^{U}$ for personalization.

First, to contextualize the predictions, we use the predicted session context information from Section~\ref{sec:method:context_prediction}. 
Specifically, for each session, we select the top-$K$\footnote{We choose $K=3$ which gives the best next-item prediction accuracy empirically.}  contexts with the highest probabilities predicted by the context predictor and concatenate their session context embedding vectors together. 
Since session contexts serve as a high-level summary of the session, the concatenation of the top-$K$ predicted context embeddings provides an accurate representation of the user's interest in the current session. 
Taking $K$ contexts instead of only one context increases the breadth of predictions and prevents erroneous predictions due to wrong context predictions from Section~\ref{sec:method:context_prediction}. 
Given a session $s$ of a user $u$ and its predicted top-$K$ contexts $\hat{C}_1, \ldots, \hat{C}_K$ (ordered by context IDs), the predicted session context representation $z_{context}$ is given as follows:
\begin{equation}
    \begin{aligned}
z_{context} =  concat(E^{C}_{\hat{C}_1}, \ldots , E^{C}_{\hat{C}_K})
\end{aligned}
\end{equation}
where $E_{c}^{C}$ is a trainable embedding  of a context $c$. All users share the same context embeddings.

Second, similar to the context predictor (Section~\ref{sec:method:context_prediction}), we summarize a user's item-level current interest within a session via a Bi-LSTM.
Given a session $s$ of a user $u$ and its observed items $i_1, \ldots , i_{k}$, the output $z_{item}$ from the item-level Bi-LSTM $\Theta_{item}$ is 
\begin{equation}
\label{eq:next_item_pred_item_level}
    \begin{aligned}
z_{item} = \Theta_{item}([E^{I}_{i_1}, \ldots , E^{I}_{i_k}]).
\end{aligned}
\end{equation}
$E_{i}^{I}$ is the item embedding of an item $i$.
Note that we should not re-use $\Theta_{short}$ and $z_{short}$ from Section~\ref{sec:method:context_prediction} in Equation~\eqref{eq:next_item_pred_item_level} since they are optimized to predict session contexts, not to predict next items.

Third, similar to the context predictor, we use a user embedding $E^{U}_{u}$ as one of the input features for the next-item predictor. A user embedding can personalize the predictions.
Furthermore, a user embedding can ease the cold-start problem when zero or only a few items are observed in a session.

Finally, \method has a concatenation layer that concatenates the above representations ($z_{context}, z_{item}, E^{U}_{u}$) and feeds them to fully connected and Softmax layers to generate next-item recommendation probabilities $\hat{p} \in \mathbb{R}^{|I| \times 1}$, where $|I|$ is the number of items. 
\begin{equation}
    \begin{split}
\hat{z}_{next-item} &= FC_{2}(concat(z_{context},z_{item},E^{U}_{u})) \\
\hat{p} &= softmax(\hat{z}_{next-item})
\end{split}
\end{equation}
We train the next-item predictor  $\Theta_{next-item} = \{\Theta_{item},E^{U},E^{I},E^{C},FC_2\}$ with the following Cross-Entropy loss and Adam optimizer on all training interactions $X_{train}$.
\begin{equation}
    \begin{aligned}
        \mathbf{L}_{next-item}(\hat{p}) = -\sum_{i=1}^{|I|} {p_i \log (\hat{p_{i}})}
    \end{aligned}
    \label{eq:next_item_prediction}
\end{equation}
where $p$ is a one-hot vector containing the ground-truth next-item ($i_{k+1}$) of the current session $s$ of the user $u$.

Using the trained $\Theta_{next-item}$, \method generates the next-item probability vectors ($\hat{p} \in \mathbb{R}^{|I| \times 1}$) for all test interactions $X_{test}$.

There are no shared model parameters or joint-training between the context predictor and next-item predictor, since they are optimized to solve different tasks. The next-item predictor only utilizes context prediction results of sessions from the context predictor.

\section{Experiments}
\label{sec:exp}
\textbf{Datasets:}
Table~\ref{tab:dataset} lists the statistics of the datasets. We created sessions of all datasets with a 1-hour idle threshold since the datasets do not include session information.
We also filter users with less than 10 interactions in all datasets.

	\begin{table}[t!]
	\footnotesize
		\caption{Summary of datasets and sessions used for experiments.}
		\begin{tabular}{| c | c | c | c | c | c |}
			\toprule
			\textbf{Name} & \textbf{Users} & \textbf{Items} & \textbf{Interactions}  & \textbf{Sessions}  & \textbf{\begin{tabular}[c]{@{}c@{}}Avg session \\ length \end{tabular}}  \\
			\midrule
			Gowalla~\cite{cho2011friendship} & 69,332 & 10,000  & 1,250,045 & 915,135  & 1.20\\
			LastFM~\cite{LastFM} & 954 & 1,000 &  258,620   & 167,382 & 1.54 \\
			Foursquare~\cite{yuan2013time} & 2,321 & 5,596  & 194,105  & 42,881 & 4.49\\
			Reddit~\cite{Reddit} & 8,640 & 966 & 134,489  & 55,698 & 2.03\\
		\bottomrule
		\end{tabular}	
		\label{tab:dataset}
	\end{table}

\begin{table}[t!]
\caption{\textit{Next-item prediction performance of \method and baselines.} 
The {\color{blue!75}most accurate model} in each column is colored {\color{blue!75}blue}, and the {\color{blue!20}second best is light blue}.
}
\begin{subtable}{0.5\textwidth}
\centering
\footnotesize
\hspace{-3mm}
\begin{tabular}{|c|c|c|c|c|}
\hline
\textbf{Model / Dataset}  & \textbf{Gowalla} & \textbf{Foursquare} & \textbf{Reddit} & \textbf{LastFM} \\ \hline
\textbf{\grurec~\cite{hidasi2018recurrent}} &  0.27724      &   0.06696      &    0.63536    &  0.12587         \\ \hline
\textbf{\tagnn~\cite{yu2020tagnn}} &    \cellcolor{blue!10} 0.32614    &    0.11189    &    0.67666     &  \cellcolor{blue!10} 0.13375          \\ \hline
\textbf{\cotrec~\cite{COTREC}} &  0.17464      &    0.11119    &   0.45382      &   0.09295       \\ \hline
\textbf{\csrm~\cite{wang2019collaborative}} &     0.31326  &     \cellcolor{blue!10} 0.12807  &    \cellcolor{blue!10} 0.68894    &  0.13190     \\ \hline
% \multicolumn{5}{|c|}{Proposed method}\\\hline
\textbf{\method (proposed)  }                                                  & \cellcolor{blue!25} \textbf{0.35975}     & \cellcolor{blue!25} \textbf{0.17483}         &   \cellcolor{blue!25} \textbf{0.72661}             & \cellcolor{blue!25} \textbf{0.13838}           \\ \hline
\end{tabular}
\captionsetup{font={small}}
\caption{Mean reciprocal rank (MRR) of \method and baselines    }
\label{tab:next_item_MRR}
\end{subtable}
\begin{subtable}{0.5\textwidth}
\hspace{-3mm}
\centering
\footnotesize
\begin{tabular}{|c|c|c|c|c|}
\hline
\textbf{Model / Dataset}  & \textbf{Gowalla} & \textbf{Foursquare} & \textbf{Reddit} & \textbf{LastFM} \\ \hline
\textbf{\grurec~\cite{hidasi2018recurrent}} &   0.47236     &    0.12920     &   0.75592     & 0.26280        \\ \hline
\textbf{\tagnn~\cite{yu2020tagnn}} &     0.49771    &  0.21256      &   0.78918      &   0.27121 \\ \hline
\textbf{\cotrec~\cite{COTREC}} &   0.39075     &   0.23728     &  0.76635      &      \cellcolor{blue!25} \textbf{0.29296}  \\ \hline
\textbf{\csrm~\cite{wang2019collaborative}} &  \cellcolor{blue!10} 0.50064    &    \cellcolor{blue!10} 0.24609     &   \cellcolor{blue!10} 0.81163      &  0.27971   \\ \hline
\textbf{\method          (proposed)   }                                         &   \cellcolor{blue!25} \textbf{0.55299}     &  \cellcolor{blue!25} \textbf{0.32973}        &       \cellcolor{blue!25} \textbf{0.86860}         &  \cellcolor{blue!10} 0.28380         \\ \hline
\end{tabular}
\captionsetup{font={small}}
\caption{Recall@10 of \method and baselines}
\label{tab:next_item_recall}
\end{subtable}
\label{tab:next_item_accuracy}
\end{table}
\noindent $\bullet$ Gowalla~\cite{cho2011friendship, ye2010location, yuan2013time} is a point-of-interest (POI) dataset collected in the US, represented as (user, location, timestamp). 

\noindent $\bullet$  LastFM~\cite{LastFM, guo2019streaming, ren2019repeatnet} includes the music playing history of users represented as (user, music, timestamp). 

\noindent $\bullet$ Foursquare~\cite{yuan2013time, ye2010location}  is a POI dataset collected from Singapore, which is represented as (user, location, timestamp). 

\noindent $\bullet$ Reddit~\cite{Reddit, JODIE} includes the posting history of users on subreddits represented as (user, subreddit, timestamp).

\noindent
\textbf{Baselines}: As a baseline, we use four state-of-the-art session-based recommender models: (1) \textbf{\grurec}~\cite{hidasi2018recurrent}: it utilizes diverse ranking-based loss functions and additional data samples for higher accuracy, (2) \textbf{\tagnn}~\cite{yu2020tagnn}: it uses a graph neural network and a target-aware attention module for prediction, (3) \textbf{\cotrec}~\cite{COTREC}: it combines self-supervised learning with graph co-training, and (4) \textbf{\csrm}~\cite{wang2019collaborative}: it contextualizes the current and neighborhood sessions with inner and outer memory encoders, respectively.

\noindent
\textbf{Experimental Setup}: Following ~\cite{hu2019sets2sets, wang2019kgat, meng2020exploring}, we use the first 90\% of interactions of each user for training, and the rest of the interactions are used for the test where the last 10\% of training interactions of each user are used as the validation set. 
The default hyperparameters of \method are set as follows (details in Appendix): the number of session contexts is 40, the number of predicted contexts per session is 3, the user and item embedding sizes are 256, and the contextual embedding size is 32.
For baselines, we use hyperparameters recommended in their original publications.
Other hyperparameters are set as follows: the maximum training epoch is set to 200, a learning rate is set to 0.001, the batch size is set to 1024, and the maximum sequence length per user is set to 50. 
We report average values of evaluation metrics measured with 5 repetitions. 
To measure statistical significance, we use a one-tailed T-test.

\subsection{Next-item Prediction Accuracy}
\label{sec:exp:next_item_prediction}
Table~\ref{tab:next_item_accuracy} shows Mean Reciprocal Rank (MRR)~\cite{voorhees1999trec} and Recall@10 metrics of \method and the state-of-the-art methods on the four datasets. Among all methods, \method mostly shows the best next-item prediction performance, \textbf{with statistical significance (p-values < 0.05)}, according to both the metrics across all four datasets. 
On the Foursquare dataset, which is the hardest-to-predict dataset for baselines, \method presents at least 36.5\% and 34.0\% improvements in MRR and Recall@10 metrics compared to the baselines.

\begin{figure}[t!]
    \centering
    \includegraphics[width=5cm]{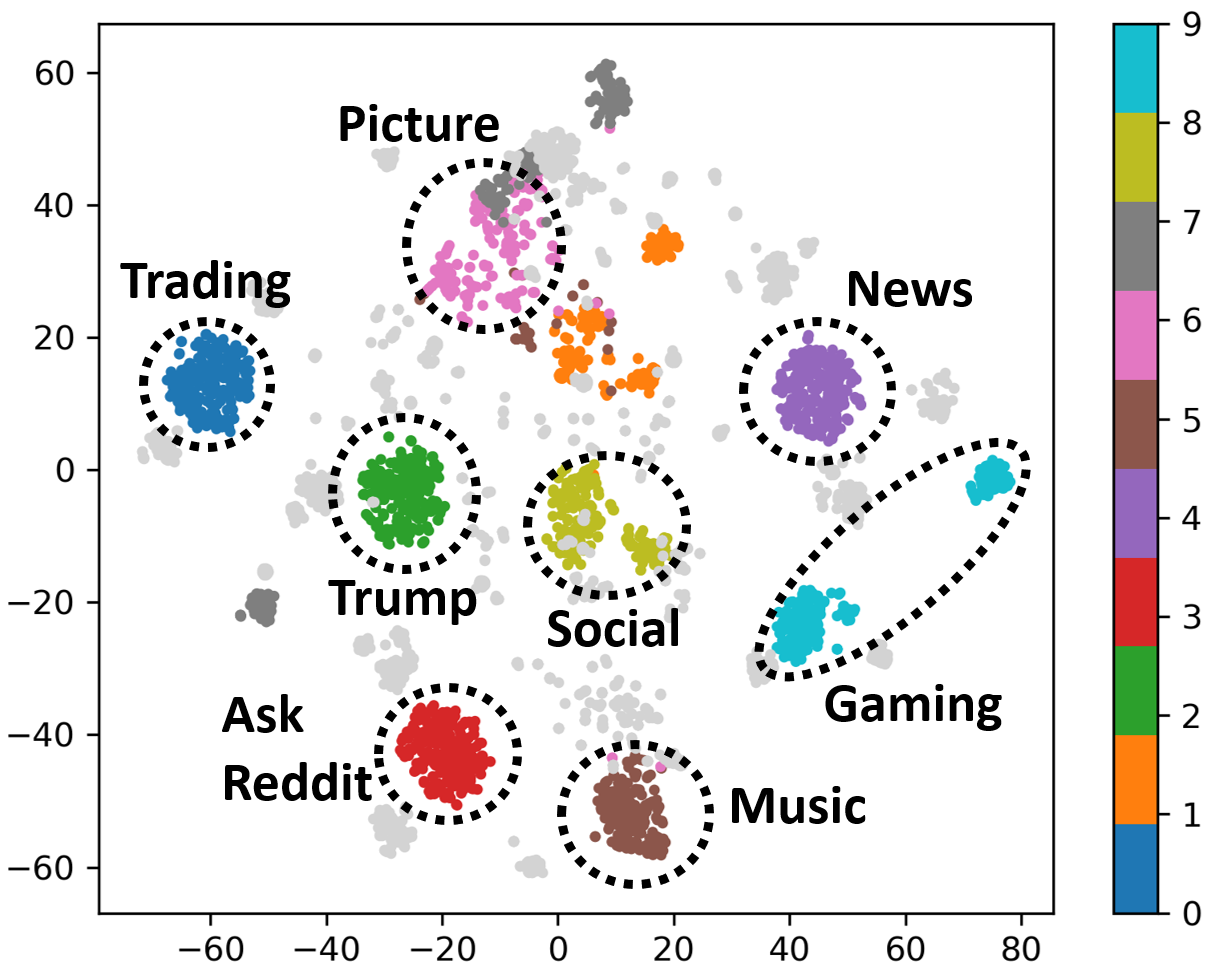}
    \caption{Implicit session context assignments by \method on the Reddit dataset. \method identifies unique contexts of session clusters. 
    }
	\label{fig:exp_case_study}
\end{figure}

\subsection{Session Contexts Evaluation}
\label{sec:exp:case_study}

\method finds the implicit contexts of sessions via clustering of session embeddings. Here, we verify the correctness of the derived session contexts. Since there is no ground-truth session context information available, we conduct a manual verification. 
We use the Reddit dataset as it includes information about the items (i.e., subreddits), such as the subreddit name, and text and title of posts (i.e., items). We first choose the top-10 clusters by size. For each cluster, we select the 100 sessions closest to the cluster center in the embedding space. After that, we analyze the items (posts) in the sessions of each cluster and manually verify if the sessions are semantically similar. If they are, we assign a context to the cluster. 

Figure~\ref{fig:exp_case_study} shows a visualization of session contexts found on the Reddit dataset, where each data point is a session, and its color represents its cluster. 
Data points with a light gray color indicate randomly sampled sessions that are not in the top-10 clusters.
We use t-SNE~\cite{van2008visualizing} to map and visualize the session embeddings to two-dimensional space. As shown in the figure, we find distinct contexts of session clusters like Trading, Music, and News topics. The dense clusters and their reasonable contextual meanings substantiate the correctness of the session contextualization approach of \method.

\section{Conclusion}\label{sec:conc}
We showed that by assigning and predicting session contexts, next-item recommendations can be improved. This work can help recommendation engines better understand their users by identifying implicit contexts. Future works include handling cold-start users and items effectively with their metadata and optimizing different neural architectures (e.g., Transformer) as the backbone of \method.
	
\bibliographystyle{ACM-Reference-Format}
\bibliography{references}


\begin{thebibliography}{45}

%%% the Web address.

\ifx \showCODEN    \undefined \def \showCODEN     #1{\unskip}     \fi
\ifx \showDOI      \undefined \def \showDOI       #1{#1}\fi
\ifx \showISBNx    \undefined \def \showISBNx     #1{\unskip}     \fi
\ifx \showISBNxiii \undefined \def \showISBNxiii  #1{\unskip}     \fi
\ifx \showISSN     \undefined \def \showISSN      #1{\unskip}     \fi
\ifx \showLCCN     \undefined \def \showLCCN      #1{\unskip}     \fi
\ifx \shownote     \undefined \def \shownote      #1{#1}          \fi
\ifx \showarticletitle \undefined \def \showarticletitle #1{#1}   \fi
\ifx \showURL      \undefined \def \showURL       {\relax}        \fi
% invisible to TeX
\providecommand\bibfield[2]{#2}
\providecommand\bibinfo[2]{#2}
\providecommand\natexlab[1]{#1}
\providecommand\showeprint[2][]{arXiv:#2}

\bibitem[Red(2020)]%
        {Reddit}
 \bibinfo{year}{2020}\natexlab{}.
\newblock \bibinfo{title}{Reddit data dump}.
\newblock \bibinfo{howpublished}{\url{http://files.pushshift.io/reddit/}}.
\newblock


\bibitem[Beutel et~al\mbox{.}(2018)]%
        {Ltcross}
\bibfield{author}{\bibinfo{person}{Alex Beutel}, \bibinfo{person}{Paul
  Covington}, \bibinfo{person}{Sagar Jain}, \bibinfo{person}{Can Xu},
  \bibinfo{person}{Jia Li}, \bibinfo{person}{Vince Gatto}, {and}
  \bibinfo{person}{Ed~H Chi}.} \bibinfo{year}{2018}\natexlab{}.
\newblock \showarticletitle{Latent cross: Making use of context in recurrent
  recommender systems}. In \bibinfo{booktitle}{\emph{Proceedings of the
  Eleventh ACM International Conference on Web Search and Data Mining}}.
  \bibinfo{pages}{46--54}.
\newblock


\bibitem[Cen et~al\mbox{.}(2020)]%
        {cen2020controllable}
\bibfield{author}{\bibinfo{person}{Yukuo Cen}, \bibinfo{person}{Jianwei Zhang},
  \bibinfo{person}{Xu Zou}, \bibinfo{person}{Chang Zhou},
  \bibinfo{person}{Hongxia Yang}, {and} \bibinfo{person}{Jie Tang}.}
  \bibinfo{year}{2020}\natexlab{}.
\newblock \showarticletitle{Controllable multi-interest framework for
  recommendation}. In \bibinfo{booktitle}{\emph{Proceedings of the 26th ACM
  SIGKDD International Conference on Knowledge Discovery \& Data Mining}}.
  \bibinfo{pages}{2942--2951}.
\newblock


\bibitem[Cho et~al\mbox{.}(2011)]%
        {cho2011friendship}
\bibfield{author}{\bibinfo{person}{Eunjoon Cho}, \bibinfo{person}{Seth~A
  Myers}, {and} \bibinfo{person}{Jure Leskovec}.}
  \bibinfo{year}{2011}\natexlab{}.
\newblock \showarticletitle{Friendship and mobility: user movement in
  location-based social networks}. In \bibinfo{booktitle}{\emph{Proceedings of
  the 17th ACM SIGKDD international conference on Knowledge discovery and data
  mining}}. \bibinfo{pages}{1082--1090}.
\newblock


\bibitem[Cho et~al\mbox{.}(2014)]%
        {cho2014properties}
\bibfield{author}{\bibinfo{person}{Kyunghyun Cho}, \bibinfo{person}{Bart van
  Merri{\"e}nboer}, \bibinfo{person}{Dzmitry Bahdanau}, {and}
  \bibinfo{person}{Yoshua Bengio}.} \bibinfo{year}{2014}\natexlab{}.
\newblock \showarticletitle{On the Properties of Neural Machine Translation:
  Encoder--Decoder Approaches}. In \bibinfo{booktitle}{\emph{Proceedings of
  SSST-8, Eighth Workshop on Syntax, Semantics and Structure in Statistical
  Translation}}. \bibinfo{pages}{103--111}.
\newblock


\bibitem[Fu et~al\mbox{.}(2018)]%
        {fu2018sequence}
\bibfield{author}{\bibinfo{person}{Hailin Fu}, \bibinfo{person}{Jianguo Li},
  \bibinfo{person}{Jiemin Chen}, \bibinfo{person}{Yong Tang}, {and}
  \bibinfo{person}{Jia Zhu}.} \bibinfo{year}{2018}\natexlab{}.
\newblock \showarticletitle{Sequence-Based Recommendation with Bidirectional
  LSTM Network}. In \bibinfo{booktitle}{\emph{Pacific Rim Conference on
  Multimedia}}. Springer, \bibinfo{pages}{428--438}.
\newblock


\bibitem[Gabriel De~Souza et~al\mbox{.}(2019)]%
        {gabriel2019contextual}
\bibfield{author}{\bibinfo{person}{P~Moreira Gabriel De~Souza},
  \bibinfo{person}{Dietmar Jannach}, {and} \bibinfo{person}{Adilson~Marques
  Da~Cunha}.} \bibinfo{year}{2019}\natexlab{}.
\newblock \showarticletitle{Contextual hybrid session-based news recommendation
  with recurrent neural networks}.
\newblock \bibinfo{journal}{\emph{IEEE Access}}  \bibinfo{volume}{7}
  (\bibinfo{year}{2019}), \bibinfo{pages}{169185--169203}.
\newblock


\bibitem[Guo et~al\mbox{.}(2019)]%
        {guo2019streaming}
\bibfield{author}{\bibinfo{person}{Lei Guo}, \bibinfo{person}{Hongzhi Yin},
  \bibinfo{person}{Qinyong Wang}, \bibinfo{person}{Tong Chen},
  \bibinfo{person}{Alexander Zhou}, {and} \bibinfo{person}{Nguyen Quoc
  Viet~Hung}.} \bibinfo{year}{2019}\natexlab{}.
\newblock \showarticletitle{Streaming session-based recommendation}. In
  \bibinfo{booktitle}{\emph{Proceedings of the 25th ACM SIGKDD International
  Conference on Knowledge Discovery \& Data Mining}}.
  \bibinfo{pages}{1569--1577}.
\newblock


\bibitem[Hamilton et~al\mbox{.}(2017)]%
        {hamilton2017inductive}
\bibfield{author}{\bibinfo{person}{William~L Hamilton}, \bibinfo{person}{Rex
  Ying}, {and} \bibinfo{person}{Jure Leskovec}.}
  \bibinfo{year}{2017}\natexlab{}.
\newblock \showarticletitle{Inductive representation learning on large graphs}.
  In \bibinfo{booktitle}{\emph{Proceedings of the 31st International Conference
  on Neural Information Processing Systems}}. \bibinfo{pages}{1025--1035}.
\newblock


\bibitem[Hansen et~al\mbox{.}(2020)]%
        {hansen2020contextual}
\bibfield{author}{\bibinfo{person}{Casper Hansen}, \bibinfo{person}{Christian
  Hansen}, \bibinfo{person}{Lucas Maystre}, \bibinfo{person}{Rishabh Mehrotra},
  \bibinfo{person}{Brian Brost}, \bibinfo{person}{Federico Tomasi}, {and}
  \bibinfo{person}{Mounia Lalmas}.} \bibinfo{year}{2020}\natexlab{}.
\newblock \showarticletitle{Contextual and sequential user embeddings for
  large-scale music recommendation}. In \bibinfo{booktitle}{\emph{Fourteenth
  ACM conference on recommender systems}}. \bibinfo{pages}{53--62}.
\newblock


\bibitem[Hidasi and Karatzoglou(2018)]%
        {hidasi2018recurrent}
\bibfield{author}{\bibinfo{person}{Bal{\'a}zs Hidasi} {and}
  \bibinfo{person}{Alexandros Karatzoglou}.} \bibinfo{year}{2018}\natexlab{}.
\newblock \showarticletitle{Recurrent neural networks with top-k gains for
  session-based recommendations}. In \bibinfo{booktitle}{\emph{Proceedings of
  the 27th ACM international conference on information and knowledge
  management}}. \bibinfo{pages}{843--852}.
\newblock


\bibitem[Hidasi and Tikk(2012)]%
        {LastFM}
\bibfield{author}{\bibinfo{person}{Bal{\'a}zs Hidasi} {and}
  \bibinfo{person}{Domonkos Tikk}.} \bibinfo{year}{2012}\natexlab{}.
\newblock \showarticletitle{Fast ALS-based tensor factorization for
  context-aware recommendation from implicit feedback}. In
  \bibinfo{booktitle}{\emph{Joint European Conference on Machine Learning and
  Knowledge Discovery in Databases}}. \bibinfo{pages}{67--82}.
\newblock


\bibitem[Hu and He(2019)]%
        {hu2019sets2sets}
\bibfield{author}{\bibinfo{person}{Haoji Hu} {and} \bibinfo{person}{Xiangnan
  He}.} \bibinfo{year}{2019}\natexlab{}.
\newblock \showarticletitle{Sets2sets: Learning from sequential sets with
  neural networks}. In \bibinfo{booktitle}{\emph{Proceedings of the 25th ACM
  SIGKDD International Conference on Knowledge Discovery \& Data Mining}}.
  \bibinfo{pages}{1491--1499}.
\newblock


\bibitem[Hu et~al\mbox{.}(2017)]%
        {hu2017diversifying}
\bibfield{author}{\bibinfo{person}{Liang Hu}, \bibinfo{person}{Longbing Cao},
  \bibinfo{person}{Shoujin Wang}, \bibinfo{person}{Guandong Xu},
  \bibinfo{person}{Jian Cao}, {and} \bibinfo{person}{Zhiping Gu}.}
  \bibinfo{year}{2017}\natexlab{}.
\newblock \showarticletitle{Diversifying Personalized Recommendation with
  User-session Context}. In \bibinfo{booktitle}{\emph{Proceedings of the
  Twenty-Sixth International Joint Conference on Artificial Intelligence}}.
  \bibinfo{pages}{1858--1864}.
\newblock


\bibitem[Kanungo et~al\mbox{.}(2002)]%
        {kanungo2002efficient}
\bibfield{author}{\bibinfo{person}{Tapas Kanungo}, \bibinfo{person}{David~M
  Mount}, \bibinfo{person}{Nathan~S Netanyahu}, \bibinfo{person}{Christine~D
  Piatko}, \bibinfo{person}{Ruth Silverman}, {and} \bibinfo{person}{Angela~Y
  Wu}.} \bibinfo{year}{2002}\natexlab{}.
\newblock \showarticletitle{An efficient k-means clustering algorithm: Analysis
  and implementation}.
\newblock \bibinfo{journal}{\emph{IEEE transactions on pattern analysis and
  machine intelligence}} \bibinfo{volume}{24}, \bibinfo{number}{7}
  (\bibinfo{year}{2002}), \bibinfo{pages}{881--892}.
\newblock


\bibitem[Kulkarni and Rodd(2020)]%
        {KULKARNI2020100255}
\bibfield{author}{\bibinfo{person}{Saurabh Kulkarni} {and}
  \bibinfo{person}{Sunil~F. Rodd}.} \bibinfo{year}{2020}\natexlab{}.
\newblock \showarticletitle{Context Aware Recommendation Systems: A review of
  the state of the art techniques}.
\newblock \bibinfo{journal}{\emph{Computer Science Review}}
  \bibinfo{volume}{37} (\bibinfo{year}{2020}), \bibinfo{pages}{100255}.
\newblock
\showISSN{1574-0137}


\bibitem[Kumar et~al\mbox{.}(2019)]%
        {JODIE}
\bibfield{author}{\bibinfo{person}{Srijan Kumar}, \bibinfo{person}{Xikun
  Zhang}, {and} \bibinfo{person}{Jure Leskovec}.}
  \bibinfo{year}{2019}\natexlab{}.
\newblock \showarticletitle{Predicting dynamic embedding trajectory in temporal
  interaction networks}. In \bibinfo{booktitle}{\emph{Proceedings of the 25th
  ACM SIGKDD international conference on knowledge discovery \& data mining}}.
  \bibinfo{pages}{1269--1278}.
\newblock


\bibitem[Li et~al\mbox{.}(2019)]%
        {li2019multi}
\bibfield{author}{\bibinfo{person}{Chao Li}, \bibinfo{person}{Zhiyuan Liu},
  \bibinfo{person}{Mengmeng Wu}, \bibinfo{person}{Yuchi Xu},
  \bibinfo{person}{Huan Zhao}, \bibinfo{person}{Pipei Huang},
  \bibinfo{person}{Guoliang Kang}, \bibinfo{person}{Qiwei Chen},
  \bibinfo{person}{Wei Li}, {and} \bibinfo{person}{Dik~Lun Lee}.}
  \bibinfo{year}{2019}\natexlab{}.
\newblock \showarticletitle{Multi-interest network with dynamic routing for
  recommendation at Tmall}. In \bibinfo{booktitle}{\emph{Proceedings of the
  28th ACM international conference on information and knowledge management}}.
  \bibinfo{pages}{2615--2623}.
\newblock


\bibitem[Li et~al\mbox{.}(2017)]%
        {li2017neural}
\bibfield{author}{\bibinfo{person}{Jing Li}, \bibinfo{person}{Pengjie Ren},
  \bibinfo{person}{Zhumin Chen}, \bibinfo{person}{Zhaochun Ren},
  \bibinfo{person}{Tao Lian}, {and} \bibinfo{person}{Jun Ma}.}
  \bibinfo{year}{2017}\natexlab{}.
\newblock \showarticletitle{Neural attentive session-based recommendation}. In
  \bibinfo{booktitle}{\emph{Proceedings of the 2017 ACM on Conference on
  Information and Knowledge Management}}. \bibinfo{pages}{1419--1428}.
\newblock


\bibitem[Liu et~al\mbox{.}(2018)]%
        {liu2018stamp}
\bibfield{author}{\bibinfo{person}{Qiao Liu}, \bibinfo{person}{Yifu Zeng},
  \bibinfo{person}{Refuoe Mokhosi}, {and} \bibinfo{person}{Haibin Zhang}.}
  \bibinfo{year}{2018}\natexlab{}.
\newblock \showarticletitle{STAMP: short-term attention/memory priority model
  for session-based recommendation}. In \bibinfo{booktitle}{\emph{Proceedings
  of the 24th ACM SIGKDD International Conference on Knowledge Discovery \&
  Data Mining}}. \bibinfo{pages}{1831--1839}.
\newblock


\bibitem[Ma et~al\mbox{.}(2020)]%
        {ma2020temporal}
\bibfield{author}{\bibinfo{person}{Yifei Ma}, \bibinfo{person}{Balakrishnan
  Narayanaswamy}, \bibinfo{person}{Haibin Lin}, {and} \bibinfo{person}{Hao
  Ding}.} \bibinfo{year}{2020}\natexlab{}.
\newblock \showarticletitle{Temporal-Contextual Recommendation in Real-Time}.
  In \bibinfo{booktitle}{\emph{Proceedings of the 26th ACM SIGKDD International
  Conference on Knowledge Discovery \& Data Mining}}.
  \bibinfo{pages}{2291--2299}.
\newblock


\bibitem[Meng et~al\mbox{.}(2020)]%
        {meng2020exploring}
\bibfield{author}{\bibinfo{person}{Zaiqiao Meng}, \bibinfo{person}{Richard
  McCreadie}, \bibinfo{person}{Craig Macdonald}, {and} \bibinfo{person}{Iadh
  Ounis}.} \bibinfo{year}{2020}\natexlab{}.
\newblock \showarticletitle{Exploring data splitting strategies for the
  evaluation of recommendation models}. In \bibinfo{booktitle}{\emph{Fourteenth
  ACM conference on recommender systems}}. \bibinfo{pages}{681--686}.
\newblock


\bibitem[Pan et~al\mbox{.}(2020)]%
        {pan2020intent}
\bibfield{author}{\bibinfo{person}{Zhiqiang Pan}, \bibinfo{person}{Fei Cai},
  \bibinfo{person}{Yanxiang Ling}, {and} \bibinfo{person}{Maarten de Rijke}.}
  \bibinfo{year}{2020}\natexlab{}.
\newblock \showarticletitle{An intent-guided collaborative machine for
  session-based recommendation}. In \bibinfo{booktitle}{\emph{Proceedings of
  the 43rd International ACM SIGIR Conference on Research and Development in
  Information Retrieval}}. \bibinfo{pages}{1833--1836}.
\newblock


\bibitem[Ren et~al\mbox{.}(2019)]%
        {ren2019repeatnet}
\bibfield{author}{\bibinfo{person}{Pengjie Ren}, \bibinfo{person}{Zhumin Chen},
  \bibinfo{person}{Jing Li}, \bibinfo{person}{Zhaochun Ren},
  \bibinfo{person}{Jun Ma}, {and} \bibinfo{person}{Maarten De~Rijke}.}
  \bibinfo{year}{2019}\natexlab{}.
\newblock \showarticletitle{Repeatnet: A repeat aware neural recommendation
  machine for session-based recommendation}. In
  \bibinfo{booktitle}{\emph{Proceedings of the AAAI Conference on Artificial
  Intelligence}}, Vol.~\bibinfo{volume}{33}. \bibinfo{pages}{4806--4813}.
\newblock


\bibitem[Schuster and Paliwal(1997)]%
        {schuster1997bidirectional}
\bibfield{author}{\bibinfo{person}{Mike Schuster} {and}
  \bibinfo{person}{Kuldip~K Paliwal}.} \bibinfo{year}{1997}\natexlab{}.
\newblock \showarticletitle{Bidirectional recurrent neural networks}.
\newblock \bibinfo{journal}{\emph{IEEE transactions on Signal Processing}}
  \bibinfo{volume}{45}, \bibinfo{number}{11} (\bibinfo{year}{1997}),
  \bibinfo{pages}{2673--2681}.
\newblock


\bibitem[Sheu and Li(2020)]%
        {sheu2020context}
\bibfield{author}{\bibinfo{person}{Heng-Shiou Sheu} {and}
  \bibinfo{person}{Sheng Li}.} \bibinfo{year}{2020}\natexlab{}.
\newblock \showarticletitle{Context-aware graph embedding for session-based
  news recommendation}. In \bibinfo{booktitle}{\emph{Fourteenth ACM conference
  on recommender systems}}. \bibinfo{pages}{657--662}.
\newblock


\bibitem[Tan et~al\mbox{.}(2021)]%
        {tan2021sparse}
\bibfield{author}{\bibinfo{person}{Qiaoyu Tan}, \bibinfo{person}{Jianwei
  Zhang}, \bibinfo{person}{Jiangchao Yao}, \bibinfo{person}{Ninghao Liu},
  \bibinfo{person}{Jingren Zhou}, \bibinfo{person}{Hongxia Yang}, {and}
  \bibinfo{person}{Xia Hu}.} \bibinfo{year}{2021}\natexlab{}.
\newblock \showarticletitle{Sparse-interest network for sequential
  recommendation}. In \bibinfo{booktitle}{\emph{Proceedings of the 14th ACM
  International Conference on Web Search and Data Mining}}.
  \bibinfo{pages}{598--606}.
\newblock


\bibitem[Twardowski(2016)]%
        {twardowski2016modelling}
\bibfield{author}{\bibinfo{person}{Bart{\l}omiej Twardowski}.}
  \bibinfo{year}{2016}\natexlab{}.
\newblock \showarticletitle{Modelling contextual information in session-aware
  recommender systems with neural networks}. In
  \bibinfo{booktitle}{\emph{Proceedings of the 10th ACM Conference on
  Recommender Systems}}. \bibinfo{pages}{273--276}.
\newblock


\bibitem[Van~der Maaten and Hinton(2008)]%
        {van2008visualizing}
\bibfield{author}{\bibinfo{person}{Laurens Van~der Maaten} {and}
  \bibinfo{person}{Geoffrey Hinton}.} \bibinfo{year}{2008}\natexlab{}.
\newblock \showarticletitle{Visualizing data using t-SNE.}
\newblock \bibinfo{journal}{\emph{Journal of machine learning research}}
  \bibinfo{volume}{9}, \bibinfo{number}{11} (\bibinfo{year}{2008}).
\newblock


\bibitem[Vaswani et~al\mbox{.}(2017)]%
        {vaswani2017attention}
\bibfield{author}{\bibinfo{person}{Ashish Vaswani}, \bibinfo{person}{Noam
  Shazeer}, \bibinfo{person}{Niki Parmar}, \bibinfo{person}{Jakob Uszkoreit},
  \bibinfo{person}{Llion Jones}, \bibinfo{person}{Aidan~N Gomez},
  \bibinfo{person}{{\L}ukasz Kaiser}, {and} \bibinfo{person}{Illia
  Polosukhin}.} \bibinfo{year}{2017}\natexlab{}.
\newblock \showarticletitle{Attention is all you need}.
\newblock \bibinfo{journal}{\emph{Advances in neural information processing
  systems}}  \bibinfo{volume}{30} (\bibinfo{year}{2017}).
\newblock


\bibitem[Voorhees et~al\mbox{.}(1999)]%
        {voorhees1999trec}
\bibfield{author}{\bibinfo{person}{Ellen~M Voorhees} {et~al\mbox{.}}}
  \bibinfo{year}{1999}\natexlab{}.
\newblock \showarticletitle{The trec-8 question answering track report.}. In
  \bibinfo{booktitle}{\emph{Text Retrieval Conference}},
  Vol.~\bibinfo{volume}{99}. \bibinfo{pages}{77--82}.
\newblock


\bibitem[Wang et~al\mbox{.}(2020b)]%
        {wang2020make}
\bibfield{author}{\bibinfo{person}{Chenyang Wang}, \bibinfo{person}{Min Zhang},
  \bibinfo{person}{Weizhi Ma}, \bibinfo{person}{Yiqun Liu}, {and}
  \bibinfo{person}{Shaoping Ma}.} \bibinfo{year}{2020}\natexlab{b}.
\newblock \showarticletitle{Make it a chorus: knowledge-and time-aware item
  modeling for sequential recommendation}. In
  \bibinfo{booktitle}{\emph{Proceedings of the 43rd International ACM SIGIR
  Conference on Research and Development in Information Retrieval}}.
  \bibinfo{pages}{109--118}.
\newblock


\bibitem[Wang et~al\mbox{.}(2019b)]%
        {wang2019collaborative}
\bibfield{author}{\bibinfo{person}{Meirui Wang}, \bibinfo{person}{Pengjie Ren},
  \bibinfo{person}{Lei Mei}, \bibinfo{person}{Zhumin Chen},
  \bibinfo{person}{Jun Ma}, {and} \bibinfo{person}{Maarten de Rijke}.}
  \bibinfo{year}{2019}\natexlab{b}.
\newblock \showarticletitle{A collaborative session-based recommendation
  approach with parallel memory modules}. In
  \bibinfo{booktitle}{\emph{Proceedings of the 42nd International ACM SIGIR
  Conference on Research and Development in Information Retrieval}}.
  \bibinfo{pages}{345--354}.
\newblock


\bibitem[Wang et~al\mbox{.}(2021)]%
        {wang2021survey}
\bibfield{author}{\bibinfo{person}{Shoujin Wang}, \bibinfo{person}{Longbing
  Cao}, \bibinfo{person}{Yan Wang}, \bibinfo{person}{Quan~Z Sheng},
  \bibinfo{person}{Mehmet~A Orgun}, {and} \bibinfo{person}{Defu Lian}.}
  \bibinfo{year}{2021}\natexlab{}.
\newblock \showarticletitle{A survey on session-based recommender systems}.
\newblock \bibinfo{journal}{\emph{ACM Computing Surveys (CSUR)}}
  \bibinfo{volume}{54}, \bibinfo{number}{7} (\bibinfo{year}{2021}),
  \bibinfo{pages}{1--38}.
\newblock


\bibitem[Wang et~al\mbox{.}(2019a)]%
        {wang2019kgat}
\bibfield{author}{\bibinfo{person}{Xiang Wang}, \bibinfo{person}{Xiangnan He},
  \bibinfo{person}{Yixin Cao}, \bibinfo{person}{Meng Liu}, {and}
  \bibinfo{person}{Tat-Seng Chua}.} \bibinfo{year}{2019}\natexlab{a}.
\newblock \showarticletitle{Kgat: Knowledge graph attention network for
  recommendation}. In \bibinfo{booktitle}{\emph{Proceedings of the 25th ACM
  SIGKDD international conference on knowledge discovery \& data mining}}.
  \bibinfo{pages}{950--958}.
\newblock


\bibitem[Wang et~al\mbox{.}(2020a)]%
        {wang2020global}
\bibfield{author}{\bibinfo{person}{Ziyang Wang}, \bibinfo{person}{Wei Wei},
  \bibinfo{person}{Gao Cong}, \bibinfo{person}{Xiao-Li Li},
  \bibinfo{person}{Xian-Ling Mao}, {and} \bibinfo{person}{Minghui Qiu}.}
  \bibinfo{year}{2020}\natexlab{a}.
\newblock \showarticletitle{Global context enhanced graph neural networks for
  session-based recommendation}. In \bibinfo{booktitle}{\emph{Proceedings of
  the 43rd International ACM SIGIR Conference on Research and Development in
  Information Retrieval}}. \bibinfo{pages}{169--178}.
\newblock


\bibitem[Wu et~al\mbox{.}(2019)]%
        {wu2019session}
\bibfield{author}{\bibinfo{person}{Shu Wu}, \bibinfo{person}{Yuyuan Tang},
  \bibinfo{person}{Yanqiao Zhu}, \bibinfo{person}{Liang Wang},
  \bibinfo{person}{Xing Xie}, {and} \bibinfo{person}{Tieniu Tan}.}
  \bibinfo{year}{2019}\natexlab{}.
\newblock \showarticletitle{Session-based recommendation with graph neural
  networks}. In \bibinfo{booktitle}{\emph{Proceedings of the AAAI Conference on
  Artificial Intelligence}}, Vol.~\bibinfo{volume}{33}.
  \bibinfo{pages}{346--353}.
\newblock


\bibitem[Xia et~al\mbox{.}(2021)]%
        {COTREC}
\bibfield{author}{\bibinfo{person}{Xin Xia}, \bibinfo{person}{Hongzhi Yin},
  \bibinfo{person}{Junliang Yu}, \bibinfo{person}{Yingxia Shao}, {and}
  \bibinfo{person}{Lizhen Cui}.} \bibinfo{year}{2021}\natexlab{}.
\newblock \showarticletitle{{Self-Supervised Graph Co-Training for
  Session-based Recommendation}}. In \bibinfo{booktitle}{\emph{30th ACM
  International Conference on Information and Knowledge Management}}.
  \bibinfo{publisher}{ACM}.
\newblock


\bibitem[Xu et~al\mbox{.}(2019)]%
        {xu2019graph}
\bibfield{author}{\bibinfo{person}{Chengfeng Xu}, \bibinfo{person}{Pengpeng
  Zhao}, \bibinfo{person}{Yanchi Liu}, \bibinfo{person}{Victor~S Sheng},
  \bibinfo{person}{Jiajie Xu}, \bibinfo{person}{Fuzhen Zhuang},
  \bibinfo{person}{Junhua Fang}, {and} \bibinfo{person}{Xiaofang Zhou}.}
  \bibinfo{year}{2019}\natexlab{}.
\newblock \showarticletitle{Graph Contextualized Self-Attention Network for
  Session-based Recommendation.}. In \bibinfo{booktitle}{\emph{Proceedings of
  the Twenty-Eighth International Joint Conference on Artificial
  Intelligence}}, Vol.~\bibinfo{volume}{19}. \bibinfo{pages}{3940--3946}.
\newblock


\bibitem[Ye et~al\mbox{.}(2010)]%
        {ye2010location}
\bibfield{author}{\bibinfo{person}{Mao Ye}, \bibinfo{person}{Peifeng Yin},
  {and} \bibinfo{person}{Wang-Chien Lee}.} \bibinfo{year}{2010}\natexlab{}.
\newblock \showarticletitle{Location recommendation for location-based social
  networks}. In \bibinfo{booktitle}{\emph{Proceedings of the 18th SIGSPATIAL
  international conference on advances in geographic information systems}}.
  \bibinfo{pages}{458--461}.
\newblock


\bibitem[You et~al\mbox{.}(2019)]%
        {you2019hierarchical}
\bibfield{author}{\bibinfo{person}{Jiaxuan You}, \bibinfo{person}{Yichen Wang},
  \bibinfo{person}{Aditya Pal}, \bibinfo{person}{Pong Eksombatchai},
  \bibinfo{person}{Chuck Rosenburg}, {and} \bibinfo{person}{Jure Leskovec}.}
  \bibinfo{year}{2019}\natexlab{}.
\newblock \showarticletitle{Hierarchical temporal convolutional networks for
  dynamic recommender systems}. In \bibinfo{booktitle}{\emph{The World Wide Web
  Conference}}. \bibinfo{pages}{2236--2246}.
\newblock


\bibitem[Yu et~al\mbox{.}(2020)]%
        {yu2020tagnn}
\bibfield{author}{\bibinfo{person}{Feng Yu}, \bibinfo{person}{Yanqiao Zhu},
  \bibinfo{person}{Qiang Liu}, \bibinfo{person}{Shu Wu}, \bibinfo{person}{Liang
  Wang}, {and} \bibinfo{person}{Tieniu Tan}.} \bibinfo{year}{2020}\natexlab{}.
\newblock \showarticletitle{TAGNN: Target attentive graph neural networks for
  session-based recommendation}. In \bibinfo{booktitle}{\emph{Proceedings of
  the 43rd International ACM SIGIR Conference on Research and Development in
  Information Retrieval}}. \bibinfo{pages}{1921--1924}.
\newblock


\bibitem[Yuan et~al\mbox{.}(2020)]%
        {yuan2020future}
\bibfield{author}{\bibinfo{person}{Fajie Yuan}, \bibinfo{person}{Xiangnan He},
  \bibinfo{person}{Haochuan Jiang}, \bibinfo{person}{Guibing Guo},
  \bibinfo{person}{Jian Xiong}, \bibinfo{person}{Zhezhao Xu}, {and}
  \bibinfo{person}{Yilin Xiong}.} \bibinfo{year}{2020}\natexlab{}.
\newblock \showarticletitle{Future data helps training: Modeling future
  contexts for session-based recommendation}. In
  \bibinfo{booktitle}{\emph{Proceedings of The Web Conference 2020}}.
  \bibinfo{pages}{303--313}.
\newblock


\bibitem[Yuan et~al\mbox{.}(2013)]%
        {yuan2013time}
\bibfield{author}{\bibinfo{person}{Quan Yuan}, \bibinfo{person}{Gao Cong},
  \bibinfo{person}{Zongyang Ma}, \bibinfo{person}{Aixin Sun}, {and}
  \bibinfo{person}{Nadia~Magnenat Thalmann}.} \bibinfo{year}{2013}\natexlab{}.
\newblock \showarticletitle{Time-aware point-of-interest recommendation}. In
  \bibinfo{booktitle}{\emph{Proceedings of the 36th international ACM SIGIR
  conference on Research and development in information retrieval}}.
  \bibinfo{pages}{363--372}.
\newblock


\bibitem[Zhao et~al\mbox{.}(2020)]%
        {zhao2020deep}
\bibfield{author}{\bibinfo{person}{Chuanchuan Zhao}, \bibinfo{person}{Jinguo
  You}, \bibinfo{person}{Xinxian Wen}, {and} \bibinfo{person}{Xiaowu Li}.}
  \bibinfo{year}{2020}\natexlab{}.
\newblock \showarticletitle{Deep Bi-LSTM Networks for Sequential
  Recommendation}.
\newblock \bibinfo{journal}{\emph{Entropy}} \bibinfo{volume}{22},
  \bibinfo{number}{8} (\bibinfo{year}{2020}), \bibinfo{pages}{870}.
\newblock


\end{thebibliography}
	
\clearpage

\section{Appendix}
\label{sec:appendix}

\subsection{Ablation Study of \method}
\label{sec:exp:ablation}
The main novelty of \method is predicting session contexts and using the predicted session contexts to enhance next-item predictions. To test the usefulness of the session contexts, we perform an ablation study that checks prediction performances of \method with and without the session contexts. 
Table~\ref{tab:ablation_contextual_embedding} shows the result.
\method uses the top-$K$ session context embeddings, along with user embeddings and item-level interest vectors, as input features to the model.
The baseline model does not use the session context embeddings as input but uses the user embeddings and item-level interest vectors. 
We observe that the next-item predictor with contextual embeddings outperforms the baseline without them across all datasets with significant differences as per both metrics (p-values < 0.05).

\subsection{Hyperparameter Sensitivity of \method}
\label{sec:appendix:hyperparameter}

We investigate how much the hyperparameters of \method affect the next-item prediction accuracy of the model. The hyperparameters include the number of session contexts, the number of predicted contexts per session, the user and item embedding size, and the contextual embedding size.
We change one hyperparameter while fixing all the others to default values offered in Section~\ref{sec:exp}.

\begin{figure}[b!]
	\centering
	\hspace{-3mm}
	\begin{subfigure}[t]{0.23\textwidth}
		\includegraphics[width=4cm]{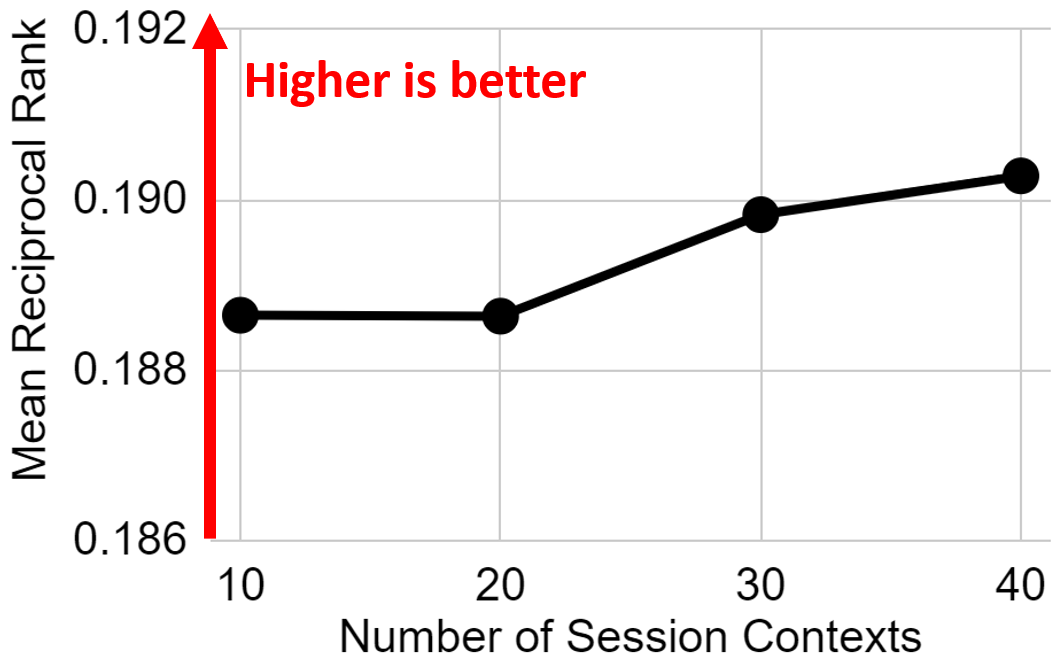}
		\captionsetup{justification=centering}
		\caption{Number of session contexts}
		\label{fig:hyper_contexts}
	\end{subfigure}
	\hspace{3mm}
	\begin{subfigure}[t]{0.23\textwidth}
		\includegraphics[width=4cm]{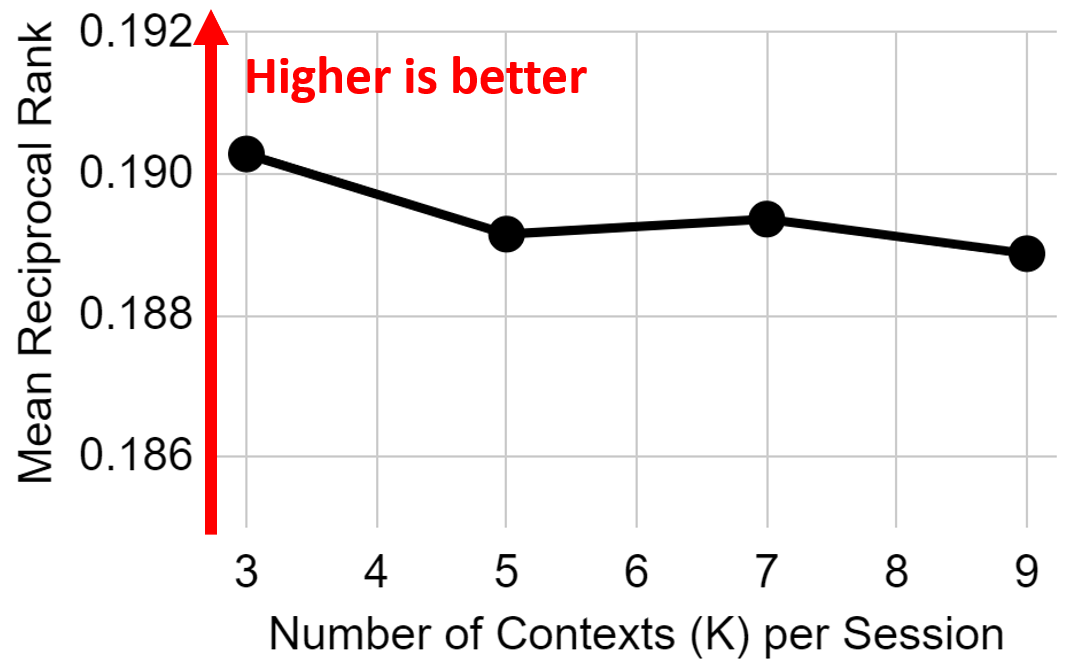}
		\captionsetup{justification=centering}
		\caption{Predicted contexts per session}
		\label{fig:hyper_topk}
	\end{subfigure}
	\\
	\hspace{-3mm}
	\begin{subfigure}[t]{0.23\textwidth}
		\includegraphics[width=4cm]{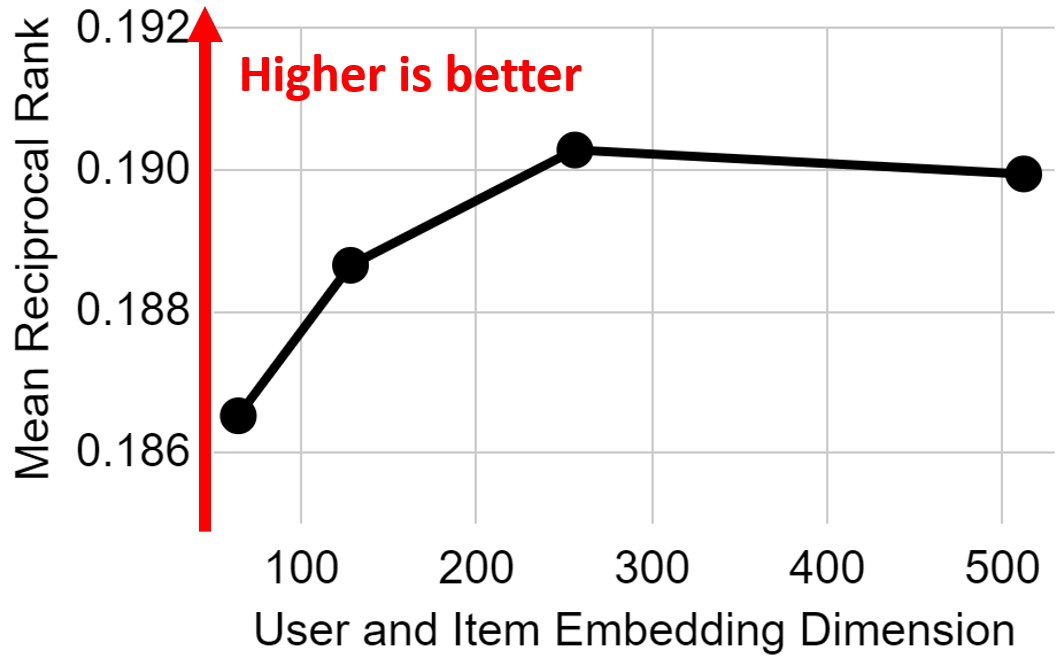}
		\captionsetup{justification=centering}
		\caption{User and item embedding dimension}
		\label{fig:hyper_user_item_emb_dim}
	\end{subfigure}
	\hspace{3mm}
	\begin{subfigure}[t]{0.23\textwidth}
		\includegraphics[width=4cm]{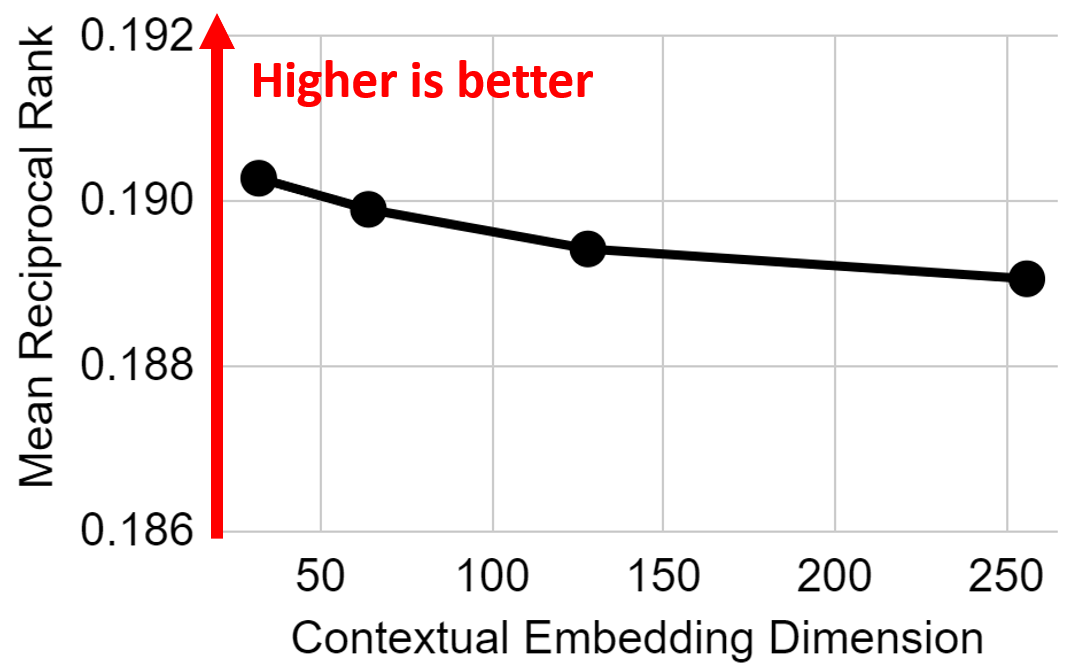}
		\captionsetup{justification=centering}
		\caption{Contextual embedding dimension}
		\label{fig:hyper_context_emb_dim}
	\end{subfigure}
	\caption{\textit{Hyperparameter sensitivity plots.}
	}
	\label{fig:hyperparameter}
\end{figure}

Figure~\ref{fig:hyperparameter} exhibits the hyperparameter sensitivity of \method with respect to the MRR metric on the Foursquare dataset (the hardest one to predict).
As the number of session contexts increases, we observe performance improvements since a larger number of contexts have more representational power to summarize the current session. On the other hand, a small number of predicted contexts per session is preferred, as additional contextual embeddings may not contain relevant information about the current session.
Regarding the embedding sizes, we find medium-size user and item embeddings are appropriate since too small or large embeddings can lead to under-fitting and over-fitting, respectively. Small-sized contextual embeddings are better than large-sized ones, since several contextual embeddings are concatenated together which provides sufficient prediction power.

\begin{table}[b!]
\centering
\footnotesize
\hspace{-3mm}
\begin{tabular}{|c|c|c|c|c|}
\hline
\textbf{Metrics / Dataset}  & \textbf{Gowalla} & \textbf{Foursquare} & \textbf{Reddit} & \textbf{LastFM} \\ \hline
\multicolumn{5}{|c|}{Next-item predictor \textit{without} session contextual embeddings    }\\\hline
\textbf{MRR}         &    0.34003     & 0.15280           &   0.70406      & 0.12328     \\ \hline
\textbf{Recall@10} 
&   0.52818       & 0.26557          &     0.83322       &  0.25438       \\ \hline
\multicolumn{5}{|c|}{\begin{tabular}[c]{@{}c@{}}Next-item predictor \textit{with} session contextual embeddings (Proposed)\end{tabular}}\\\hline
\textbf{MRR}                                                    & \cellcolor{blue!25} \textbf{0.35975}     & \cellcolor{blue!25} \textbf{0.17483}         &   \cellcolor{blue!25} \textbf{0.72661}             & \cellcolor{blue!25} \textbf{0.13838}            \\ \hline
\textbf{Recall@10}                                                              &   \cellcolor{blue!25} \textbf{0.55299}     &  \cellcolor{blue!25} \textbf{0.32973}        &       \cellcolor{blue!25} \textbf{0.86860}         &  \cellcolor{blue!25} \textbf{0.28380}         \\ \hline
\end{tabular}
\caption{\textit{Ablation study of the session context embeddings of \method.} 
}
\label{tab:ablation_contextual_embedding}
\end{table}

	\section*{Acknowledgments}
	{
	This research is supported in part by Georgia Institute of Technology, IDEaS, Adobe, and Microsoft Azure. S.O. was partly supported by ML@GT, Twitch, and Kwanjeong fellowships. We thank the reviewers for their feedback.
	}	
\end{document}